\documentclass{article}

\usepackage{microtype}
\usepackage{graphicx}
\usepackage{subfigure}
\usepackage{booktabs} 
\usepackage{csquotes}
\usepackage{tabularx}
\usepackage{natbib}
\usepackage{placeins}
\DeclareUnicodeCharacter{2060}{\nolinebreak}

\usepackage{xurl}

\usepackage[accepted]{icml2024}

\usepackage{amsmath}
\usepackage{amssymb}
\usepackage{mathtools}
\usepackage{amsthm}

\usepackage[capitalize,noabbrev]{cleveref}

\theoremstyle{plain}

\theoremstyle{definition}

\theoremstyle{remark}

\usepackage[textsize=tiny]{todonotes}

\icmltitlerunning{Position: Technical Research and Talent is Needed for Effective AI Governance}
\begin{document}

\twocolumn[
\icmltitle{Position: Technical Research and Talent is Needed for Effective AI Governance}



\icmlsetsymbol{equal}{*}

\begin{icmlauthorlist}
\icmlauthor{Anka Reuel}{equal,yyy}
\icmlauthor{Lisa Soder}{equal,int}
\icmlauthor{Ben Bucknall}{govai,aigi}
\icmlauthor{Trond Arne Undheim}{yyy}
\end{icmlauthorlist}

\icmlaffiliation{yyy}{Department of Computer Science, Stanford University, Stanford, US}
\icmlaffiliation{govai}{Centre for the Governance of AI, Oxford, UK}
\icmlaffiliation{aigi}{Oxford Martin AI Governance Initiative, Oxford, UK}
\icmlaffiliation{int}{London School of Economics, London, UK}

\icmlcorrespondingauthor{Anka Reuel}{anka@cs.stanford.edu}



\icmlkeywords{Machine Learning, ICML, AI Governance}

\vskip 0.3in
]



\printAffiliationsAndNotice{\icmlEqualContribution} 


\begin{abstract}
In light of recent advancements in AI capabilities and the increasingly widespread integration of AI systems into society, governments worldwide are actively seeking to mitigate the potential harms and risks associated with these technologies through regulation and other governance tools. However, there exist significant gaps between governance aspirations and the current state of the technical tooling necessary for their realisation. In this position paper, we survey policy documents published by public-sector institutions in the EU, US, and China to highlight specific areas of disconnect between the technical requirements necessary for enacting proposed policy actions, and the current technical state of the art. Our analysis motivates a call for tighter integration of the AI/ML research community within AI governance in order to i) catalyse technical research aimed at bridging the gap between current and supposed technical underpinnings of regulatory action, as well as ii) increase the level of technical expertise within governing institutions so as to inform and guide effective governance of AI.
\end{abstract}

\section{Introduction}
\label{introduction}

The growing integration of artificial intelligence (AI) into various aspects of society over the past decade, including education, hiring, and finance, has became particularly noticeable with ChatGPT's release in late 2022. This event significantly heightened AI's visibility among both the general public and policymakers. The adoption of the technology, especially generative AI, has since surged across various sectors and is projected to further accelerate workforce automation and occupational switches \cite{mckinsey2023futurework}. As the adoption of AI grows, so too does the awareness of its potential harms and risks. Over the past year, there has been a notable rise in reported AI-related incidents \cite{maslej2024index}—i.e., instances where the failure or misuse of an AI system led to real-world harm, such as the production of harmful deepfakes or privacy breaches \cite{mcgregor2021preventing}. Given these developments, the need for effective, robust, and adaptable AI governance at national and international levels is more pressing than ever \cite{trager2023international}. 

Governments around the world are responding to these challenges. For instance, Stanford's 2024 AI Index reveals a surge in AI-related laws, with relevant legislation increasing from one instance in 2016 to 28 by 2023 across 128 countries \cite{maslej2024index}. Among these developments, for example, China has recently enacted a new regulation on generative AI \cite{generative2023chinalaw}, while the EU has introduced the AI Act, with some rules scheduled to come into effect as early as Summer 2024 \cite{European_Commission2023-hs}. Likewise, the US has issued an Executive Order (E.O.) on the ``Safe, Secure, and Trustworthy Development and Use of Artificial Intelligence'' and garnered voluntary commitments from major AI labs for governance measures \cite{whitehouse2023voluntary}.

\begin{figure*}[bt]
    \centering
    \includegraphics[width=\textwidth]{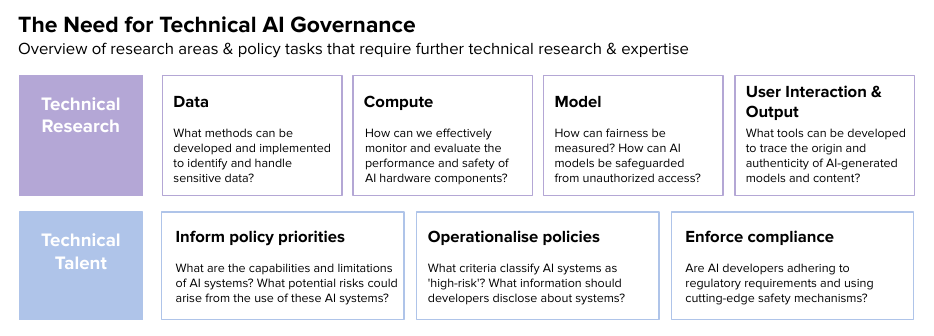} 
    \caption{Overview of the role of technical research and expertise in AI governance. We propose that (i) targeted technical research is needed to address open tooling and research questions around data, compute, models, and deployment, and (ii) more technical expertise within governing institutions will be required to inform policy priorities, operationalise them into concrete policies, and enforce them effectively.}
    \label{fig:icml1}
\end{figure*}

However, current gaps—both in terms of the technical tools and talent available to enact policy objectives—might negatively impact the capacity of governance actors to mitigate risks through regulation, potentially even lead to counterproductive outcomes \cite{guha2023ai}. Poorly designed and executed policies, such as mandatory registration of AI systems, information disclosure requirements, or pre-market approvals, could not only fail to ensure safe and reliable products but also burden AI research. Moreover, such policies risk creating entry barriers that disproportionately favor well-established, larger AI companies, potentially limiting competition and stifling innovation within the AI field \cite{dal2006capture, Wu2023capture}. To counter such issues, \textbf{we advocate for the following pair of positions:}

\begin{itemize}
    \item \textbf{Targeted AI/ML research is urgently needed to ensure the effective enactment of current regulatory proposals.}
    \item \textbf{Closer collaboration between technical researchers and policy makers is necessary to ensure informed and effective governance.}
\end{itemize}

While advocating for these claims, we acknowledge that the governance of AI inevitably involves complex, value-laden decisions \cite{gordon2022mapping, kaminski2023regulating}, and reject techno-solutionist approaches to mitigating harms from AI. Addressing the manifold issues presented by such a technology necessitates a holistic and inter-disciplinary approach combining judicial, political, technological and broader societal insights, and we are consequently arguing for a closer collaboration between researchers and practitioners in these diverse and complementary fields.

This paper proceeds as follows. Section ~\ref{regtexts} presents a systematic review of policy documents published in the US, the EU, and China respectively, identifying the technical assumptions and tools upon which they rely.
In Section ~\ref{techrequ}, these requirements are contrasted against the current state of the art (SOTA) in AI/ML research, aiming to drawing attention to disparities between policy aspirations and technical realities. Section ~\ref{researchpriorities} suggests high-level research areas where further work could aid in addressing these gaps. Section ~\ref{talentgap} discusses the challenge posed by shortages of technical talent in governance institutions (Section ~\ref{talentshortage}) before highlighting opportunities for the AI/ML research community to assist in ongoing governance efforts (Section ~\ref{researchersupport}). Section ~\ref{conclusion} concludes.

\section{Gap Between Policy Aspirations and Technical Research}\label{researchgap}

Recent policy developments depend to a good extent on technical solutions that don't necessarily exist (yet). Often, policymakers set forth regulatory frameworks with the intention of ensuring, for instance, the safety, non-discrimination, and transparency in AI systems \cite{EU2024AIA}. However, these regulations sometimes depend on technological solutions that are not yet feasible given the current SOTA in AI/ML research. This mismatch implies that some regulatory requirements might be  unattainable, highlighting a gap between regulatory expectations and technological capabilities. The gap necessitates further technical research to develop the tools and methodologies needed to meet and enforce these regulatory requirements. 


In this chapter, we are providing concrete examples of this phenomenon from governance texts from the EU, US, and China, illustrating how certain policy prescriptions are outpacing the existing AI/ML research landscape, thereby emphasising the need for continued advancement and targeted, governance-supporting AI/ML research to bridge this divide. Finally, we provide an initial overview of a body of AI/ML research that the AI/ML community should prioritise to close the gap outlined above.

\subsection{Relevant Regulatory Texts}\label{regtexts}


For the purpose of this analysis, we will concentrate on AI legislation in the EU, US, and China. While AI governance frameworks proliferated rapidly across the globe in 2023, with 28 AI-related bills across 15 countries passing into law \cite{maslej2023artificial}, these three regions hold particular significance in the AI governance landscape. This is due to their influential roles shaping in AI research and development, as well as the potential for policy diffusion \cite{shipan2008mechanisms, bradford2012brussels, siegmann2022brussels}. This subsection provides an overview of the documents we're examining from each jurisdiction. The subsequent subchapter will highlight specific requirements from these texts, pinpointing areas where they diverge from SOTA AI/ML research.

\subsubsection{European Union}\label{regtexts:EU}


Our focus of analysis in the EU is on the \textit{European AI Act}, which is one of the first comprehensive pieces of legislation in this region directly focused on the regulation of AI systems \cite{EU2024AIA}. The act, which was first proposed by the EU Commission in 2021 \cite{eucommission2022aiapproach} and finally approved by the council in May 2024, is part of a broader legislative environment in the European Union, which includes the Digital Services Act, the Digital Markets Act, the General Data Protection Regulation, as well as national legislation of the member states \cite{hacker2024regulating, bogucki2022ai}. 

The \textbf{EU AI Act} \cite{EU2024AIA} adopts a horizontal regulatory approach, meaning it establishes a common set of rules that apply uniformly across different sectors or industries. It follows a 'risk-based' approach, categorising AI systems into different levels of risk, each associated with a different set of regulatory obligations. High-risk systems, such as those used in critical infrastructures or for socio-economic decisions (e.g., hiring, education, financial services), face stringent requirements like conformity assessments, transparency obligations, and detailed record-keeping, while minimal-risk systems are largely free from regulatory obligations. GPAI systems have additional technical documentation requirements, and those posing systemic risks (e.g., models trained using compute greater than $10^{25}$ FLOPs) must conduct model evaluations, adversarial testing, and report serious incidents.


\subsubsection{United States}\label{regtexts:US}

In the context of US AI governance, we focus on two key documents: the \textit{Blueprint for an AI Bill of Rights} \cite{whitehouse2022AIBoR} and \textit{Executive Order 14110} \cite{whitehouse2023EO14110}. The \textbf{Blueprint for an AI Bill of Rights} (AIBoR), released on October 4, 2022, establishes a rights-based framework that outlines principles to guide the design, development, and deployment of automated systems \cite{whitehouse2022AIBoR}. While the AIBoR is neither binding nor equips regulatory agencies with more enforcement power, we included it in our analysis as it provides a first indication of the approach in the Biden-Harris administration and already lays out resources and best practices to enact these principles.


Second, \textbf{E.O. 14110}, titled ``Executive Order on Safe, Secure, and Trustworthy Development and Use of Artificial Intelligence'' \cite{whitehouse2023EO14110} and signed by President Biden on October 30, 2023, represents an initial approach by the US government to AI regulation. It mandates several federal agencies to appoint Chief AI Officers and outlines specific responsibilities for agencies like the Department of Homeland Security, the Department of Veterans Affairs, and the National Institute of Standards and Technology \cite{whitehouse2023EO14110}. The order has been described as the most comprehensive piece of governance by the US government pertaining to AI, following earlier initiatives such as the BAIBoR \cite{ryan2023threeEO}.

\subsubsection{China}\label{regtexts:China}

For China, we look at two different pieces of regulations: the \textit{Provisions on the Management of Algorithmic Recommendations in Internet Information Services} \cite{chinalawtranslate2021recommendation} and the \textit{Provisions on the Administration of Deep Synthesis Internet Information Services} \cite{chinalawtranslate2022deepsynthesis}\footnote{Given the limited scope of the paper, we omitted the \textit{Interim Measures for the Management of Generative Artificial Intelligence Services} \cite{chinalawtranslate2023generativeai}}.

The \textbf{Provisions on the Management of Algorithmic Recommendations in Internet Information Services} \cite{chinalawtranslate2021recommendation}, introduced by China in 2021, is a regulatory framework designed to govern the use of algorithmic recommendation systems by internet service providers within the country. This legislation aims to ensure that algorithms promote positive content and uphold core socialist values, by requiring transparency in the operations of these algorithms and holding companies accountable for the content they recommend.

The \textbf{Provisions on the Administration of Deep Synthesis Internet Information Services} \cite{chinalawtranslate2022deepsynthesis}, issued by China in 2022, represents a significant regulatory step towards overseeing what the Chinese call deep synthesis technologies, including deepfakes and other AI-generated content. This regulation aims to address the challenges and risks associated with the rapid development and application of deep synthesis in internet information services, particularly the potential for spreading misinformation, violating personal rights, and undermining social stability \cite{chinalawtranslate2022deepsynthesis}.

\subsection{Technical Requirements in AI Regulations}\label{techrequ}

To structure the analysis of the documents, we focus on technical requirements across four areas where research and regulatory requirements intersect, based on a categorisation by \cite{bommasani2023compliance}: Data, Compute, Model, Output. Across these four categories, many of the techniques and tools required to effectively implement the proposed policies are either underdeveloped or non-existent. This disparity between regulatory requirements and the current state-of-the-art in AI/ML research poses significant challenges to the effective governance of AI systems. The following section illuminates specific examples of these gaps and their implications for AI governance.


\subsubsection{Data}\label{req:data}

All three regions, the EU, the US, and China, have data governance requirements in the regulatory documents outlined in Section~\ref{regtexts}. For example, the EU AI Act states:

\begin{displayquote}
    \textit{``The data sets should also have the appropriate statistical properties, including as regards the persons or groups of persons in relation to whom the high-risk AI system is intended to be used, with specific attention to the mitigation of possible biases in the data sets.''} \citep[Recital 67]{EU2024AIA}
\end{displayquote}

Technical bias identification and mitigation within AI systems comes with several challenges: For example, both necessitate a clear definition of fairness -- a concept that remains inherently ambiguous and subjective in the context of AI \cite{mehrabi2021survey, pagano2023bias}. Despite the critical importance of ensuring equitable outcomes in AI applications, there is currently no universally accepted framework or structured approach for determining the appropriate technical implementation of fairness.\\

In the US, the E.O.and the BAIBoR both outline requirements for data governance and data privacy:

\begin{displayquote}
    \textit{``Artificial Intelligence's capabilities [...] can increase the risk that personal data could be exploited and exposed. To combat this risk, the Federal Government will ensure that the collection, use, and retention of data is lawful, is secure, and mitigates privacy and confidentiality risks. Agencies shall use available policy and technical tools, including privacy-enhancing technologies (PETs) where appropriate, to protect privacy [...]''} \citep[Section 2(f)]{whitehouse2023EO14110}
\end{displayquote}

A similar requirement is put forth in the Chinese AI regulations:

\begin{displayquote}
    \textit{``Deep synthesis service providers shall implement primary responsibility for information security, establishing and completing management systems such as for [...] data security, personal information protection [...] and shall have safe and controllable technical safeguard measures.''} \citep[Article 7]{chinalawtranslate2022deepsynthesis}
\end{displayquote}

The possibility for protection and guarantee of privacy vary depending on the type of AI system in use. For instance, LLMs have been found vulnerable to revealing personally identifiable information (PII) under adversarial attacks, such as through adversarial prompting strategies \cite{nasr2023scalable, ippolito2023preventing}. Currently, there is no robust method to safeguard privacy in these cases. Similar privacy breaches have also been observed in non-generative models \cite{rigaki2023survey, cristofaro2020privacy}.

\begin{displayquote}
    \textit{``You should be protected from abusive data practices via built-in protections and you should have agency over how data about you is used.''} \cite{whitehouse2022AIBoR}
\end{displayquote}

For users to have agency, they need to be informed about if and how their data was utilised in a model. The provenance of training data in AI systems, particularly in LLMs, is often undisclosed \cite{longpre2023data, ojewale2024ai}. At the same time, there are only limited possibilities to find out, without white-box access to the model, what was in the training data, making it difficult for users to know if and for what their data was used if it has not been proactively been disclosed by the model provider \cite{longpre2023data, longpre2024data, casper2024blackbox}. Moreover, attributing training data is difficult, especially without access to the model, and recent advancements in the field aren't applicable to all model classes \cite{park2023trak}. Therefore, even if users are aware that their data was included in the training set, they rarely understand its influence on the model's behaviour, which further limits their agency.

\subsubsection{Compute}\label{req:compute}

Both the US and the EU have registration requirements for model developers that train large models; the EU defined a corresponding FLOPs threshold beyond which a model is considered to have systemic risk (and hence is subject to more scrutiny):

\begin{displayquote}
    \textit{``A general-purpose AI model shall be presumed to have high impact capabilities pursuant to paragraph 1, point (a) [definition of systemic risk], when the cumulative amount of computation used for its training measured in floating point operations is greater than $10(^25)$.''}[Art 51]\cite{EU2024AIA}
\end{displayquote}

\begin{displayquote}
    \textit{``Where a general-purpose AI model meets the condition referred to in Article 51(1), point (a), [referring to GPAI systems with systemic risks] the relevant provider shall notify the Commission without delay [...]''} 
 \citep[Article 52, 1]{EU2024AIA}
\end{displayquote}

A related requirement can be found in the US E.O., that requires
\begin{displayquote}
    \textit{``Companies developing or demonstrating an intent to develop potential dual-use foundation models to provide the Federal Government, on an ongoing basis, with information, reports, or records regarding [...] (A) any ongoing or planned activities related to training, developing, or producing dual-use foundation models[...]''} \citep[Sec. 4.2(i)]{whitehouse2023EO14110}
\end{displayquote}


Although this requirement is not technically impossible to fulfil, enforcing it will prove challenging. Currently, there's no method to verify if a large model is being trained. Some proposed solutions include some form of on-chip governance mechanisms and chip control, e.g., using chips that either limit or report the number of computations they perform \cite{aarne2024governablechips}.

\subsubsection{Models}\label{req:model}

All regions we analysed put forward requirements for evaluating, reporting, and mitigating risks from AI systems. For example, in the EU AI Act requires

\begin{displayquote}
    \textit{``providers of general-purpose AI models with systemic risk shall: (a) perform model evaluation in accordance with standardised protocols and tools reflecting the state of the art, including conducting and documenting adversarial testing of the model with a view to identifying and mitigating systemic risks.''} \citep[Article 55]{EU2024AIA}
\end{displayquote}

Whereas the E.O.in the US has similar evaluation requirements for 'dual-use foundation models':

\begin{displayquote} 
    \textit{``(i) Companies developing or demonstrating an intent to develop potential dual-use foundation models to provide the Federal Government, on an ongoing basis, with information, reports, or records regarding the following: [...]
    (C) the results of any developed dual-use foundation model's performance in relevant AI red-team testing based on guidance developed by NIST pursuant to subsection 4.1(a)(ii) of this section, and a description of any associated measures the company has taken to meet safety objectives, such as mitigations to improve performance on these red-team tests and strengthen overall model security.''} \citep[Section 4.2 (C)]{whitehouse2023EO14110}
\end{displayquote}

Although numerous jurisdictions mandate capability evaluations across a variety of risk areas, from a technical standpoint, there is a lack of clarity on how to perform these assessments both comprehensively and reliably \cite{chang2023survey, zhou2023don}. \citet{weidinger2023sociotechnical}, in their review of current safety evaluation practices, noted that for a wide range of risk scenarios, evaluations simply do not exist yet. Even for risk areas where evaluations are available, there is still a substantial degree of uncertainty regarding their ability to accurately capture and assess specific concepts \cite{raji2021ai, liao2023hci}, and as such how much trust to put into evaluations \cite{liao2021we, zhou2023don, mitchell2023we}.

Furthermore, the US included access and security requirements for model weights in their E.O.:

\begin{displayquote}
    \textit{``(i) Companies developing or demonstrating an intent to develop potential dual-use foundation models to provide the Federal Government, on an ongoing basis, with information, reports, or records regarding the following: [...]  (B)  the ownership and possession of the model weights of any dual-use foundation models, and the physical and cybersecurity measures taken to protect those model weights''} \citep[Section 4.2 (B)]{whitehouse2023EO14110}
\end{displayquote}   

Securing models from leakage has proven to be difficult \cite{franzen2024mistralleak, vincent2023metaleak}. Some technical ideas to de-incentivise leackage of models has been to watermark models so that it can be traced back to who released a model \cite{mehta2022aime, boenisch2021systematic}. Alternative ideas included self-destructing models \cite{henderson2023self}, but both are underexplored and non-production-ready ideas that would need to be investigated further.

\subsubsection{User interaction and Output}\label{req:deployment}



Finally, the US mentions consumer protection from harmful, ineffective models:

\begin{displayquote}
    \textit{``You should be protected from unsafe or ineffective systems.''} \cite{whitehouse2022AIBoR}
\end{displayquote}

Protection from ineffective or unsafe systems is difficult to ensure. One issue is that no robust evaluation methods exist to comprehensively understand whether a model is harmless or not (see Section~\ref{req:model}). Furthermore, AI systems that are released may be ineffective. For example, in the context of generative AI, no robust technical approach to outputting hallucination-free, reliable information exists yet, which renders generative AI systems ineffective for many use cases \cite{rawte2023survey}.\\ 

Similarly, regulation from China stipulates that the output of a system has to be controlled:

\begin{displayquote}
 \textit{[...] persist [AI systems] in being oriented towards mainstream values, optimize mechanisms for algorithmic recommendation services, actively transmit positive energy, and promote the uplifting use of algorithms. The providers of algorithmic recommendation services [...] shall employ measures to prevent and stop the transmission of negative information} \cite{chinalawtranslate2021recommendation}
\end{displayquote}

While approaches such as Constitutional AI \cite{bai2022constitutional} exist to somewhat align models with certain values, these approaches are far from perfect and do not ensure adherence to the predetermined values all the time \cite{ji2023ai}. Additionally, it has been shown that these guardrails can be circumvented \cite{wolf2023fundamental, zhang2023safety}.

To reduce risks posed by synthetic content, the US E.O. lays out priorities to identify current practices, as well as support the potential development of
\begin{displayquote}
    \textit{``further science-backed standards and techniques, for (i)    authenticating content and tracking its provenance; (ii)   labeling synthetic content, such as using watermarking; (iii)  detecting synthetic content;''} \citep[Section 4.5 (a)]{whitehouse2023eocomment}
\end{displayquote}

While there have been significant advancements in watermarking AI-generated content \cite{zhao2023provable, sadasivan2023can, ghosal2023survey, frohling2021feature}, the detection of non-watermarked, AI-generated content has proven to be difficult for all modalities \cite{li2023generalizable, weber2023testing, heidari2023deepfake}. Especially in light of advancements in generated video, image, and audio quality, detection of AI-generated content will remain challenging, necessitating more research in this area to enact these regulations.

\subsection{Prioritising AI/ML Research for AI Governance}\label{researchpriorities}

\begin{figure}[htbp!]
    \centering
    \includegraphics[width=\columnwidth]{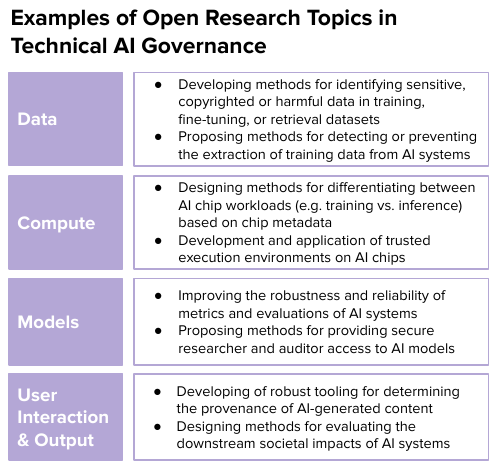} 
    \caption{Technical Research: Examples of areas that require further technical research to allow for supporting effective AI governance.}
    \label{fig:icml2}
\end{figure}

We define technical research for AI governance as 
\textit{the development and deployment of technical tools, methods, and techniques for supporting effective AI governance, as well as analysis that motivates and informs their implementation.}
Based on the comparative analysis and the gaps identified, as well as following from our definition, we suggest to prioritise AI/ML research addressing technical shortcomings in the four categories of \textit{data}, \textit{compute}, \textit{models}, and \textit{deployment} presented above. Figure \ref{fig:icml2} provides example high-level topics within each of these categories that could be addressed by future research.
These are not meant to be exhaustive lists. We also want to note that for most of the examples highlighted, there already exists research to a varying degree. However, none of the areas are sufficiently well studied, as outlined in Section~\ref{techrequ}, to close the gaps highlighted in Section~\ref{techrequ}, and hence warrant more attention from the AI/ML research community.


\section{AI Governance Requires More Technical Expertise}\label{talentgap}

A core question at the heart of any regulatory regime -- and particularly in those as complex and rapidly evolving as the AI domain -- is whether the regulator is equipped with the expertise and capacity necessary \citep{weissinger2022complexityreg, baldwin2011goodregulation} to i) inform and guide governance priorities, ii) operationalise these priorities into concrete policies and verifiable requirements and iii) enforce them. Building on the previous sections' focus on the technical tools needed to enact policies, we now turn our attention to the expertise required to develop an effective AI accountability infrastructure and governance capacity. In this section, we will first highlight some of the technical talent shortages AI governance faces. We will then elaborate on the need for AI/ML researchers to support the key tasks along the policy cycle. Finally, we will provide a brief overview and examples of avenues for ML researchers to take a more active role in AI governance efforts.

\subsection{Technical Talent Gap in AI Governance}\label{talentshortage}

Currently, there is little empirical evidence on the preparedness of governments to effectively govern AI technologies. However, graduation data in the US paints a concerning picture, with less than 1\% of AI PhD graduates choosing government careers post-study, while over 65\% enter the private sector and more than 25\% pursue academic careers \cite{zweben2017survey}. Drawing a similar conclusion, the need for more technical experts in AI governance has been raised as a key concern and priority \cite{guha2023ai, schmidt2021national, engstrom2020government}. This challenge is not unique to the US; similar worries about attracting sufficient technical talent have been voiced in other jurisdictions, including the UK \cite{aitken2022common} and EU member countries like Germany \cite{engler2022institutionalizing}.

To address some of these capacity constraints, regulators often outsource some regulatory activities to external, private sector efforts, i.e., ``regulatory intermediaries'' \cite{abbott2017introducing} such as standard-setting bodies and third-party auditors. While such an approach can offer more flexibility - for instance, private organisations such as AI auditors might be able to offer higher salaries than traditional government roles, these approaches, too, suffer from their own capacity shortages. For instance, at present, the ecosystem of third-party research organisations equipped to perform capability evaluations or risk assessments is still underdeveloped \cite{engler2023key}. Additionally, standard-setting bodies, while able to harness technical expertise from industry \citep{blind2023standards, ramanna2015political}, frequently fall short in integrating a broader and more independent range of expertise to tackle the socio-technical nature of their work \cite{edwards2022regulating}.

\subsection{The Need for Technical Expertise in AI Governance}\label{researchersupport}

In this final chapter, we highlight three different tasks across the policy lifecycle where technical researchers can contribute to supporting AI governance efforts. 

\begin{figure}[htbp!]
    \centering
    \includegraphics[width=\columnwidth]{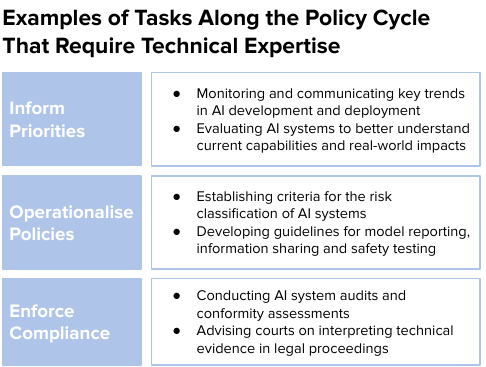} 
    \caption{Technical Talent: Examples of tasks along key steps in the policy cycle that will require technical expertise.}
    \label{fig:icml3}
\end{figure}

\subsubsection{Inform and Guide Policy Priorities}
A key challenge for AI governance is to set the right priorities and design targeted policies that neither fall short in their mandate to protect the public interest nor become excessively intrusive and hamper innovation. To achieve this goal, policymakers will need to have the necessary means to source reliable information about the progress of AI systems, their limitations, and potential future trajectories \cite{whittlestone2021governments, Bengio2024risk, clarck2024information}. 


Monitoring developments and providing scientific insights are crucial for informing policy priorities across sectors. In climate change, the Intergovernmental Panel on Climate Change informs government decisions by providing climate models and synthesizing research findings to guide policies on on emissions reductions. Similarly, central banks rely on economic data and analysis from economists and researchers to assess the state of the economy and determine appropriate actions, such as setting interest rates or adjusting the money supply. 

One such example within the AI governance space, where we've already observed a government body aiming to build the infrastructure to deliver such information, is the UK AI Safety Institute (AISI). The UK's AISI was tasked with building \textit{``a body of evidence on the risks from advanced AI [...] to lay the foundations for technically grounded international governance.''} \cite{Dsit2023-ot}. In working towards this aim, the AISI intends to carry out three main activities: developing and conducting evaluations on advanced AI systems, driving foundational AI safety research (including \textit{``building products for AI governance''} \cite{Dsit2023-ot}), and facilitating information exchange. Several other jurisdictions, including the US \cite{Nist2023-se}, Japan \cite{shimbun2023japanaisafetyinstitute} and Canada \cite{Canada2024AISI}, have followed suit, announcing initiatives akin to the AISI. 

\subsubsection{Operationalise policies}
Once policy priorities and interventions are formulated, a key challenge lies in translating these into more concrete obligations and verifiable claims. This step can encompass tasks such as defining the scope and character of activities to be integrated into risk management processes \cite{ozlati2017standards, schuett2023risk, pouget2024ai}, outlining reporting requirements \cite{kolt2024responsible, bommasani2024transparency} or establishing the criteria of what constitutes an AI system of ``systemic'' or ``high'' risk \cite{edwards2022regulating, moes2023tiers}. 

Defining different risk levels, for example, will be a key factor, both in the US and EU, in determining the level of scrutiny an AI system should undergo \cite{whitehouse2023EO14110, eucommission2022aiapproach}. However, establishing these demands a deep understanding of the current capabilities and limitations of AI systems. A lack thereof can have significant consequences, as regulatory failures in other industries have previously demonstrated. For instance, in the case of the Boeing 737 groundings, while multiple factors contributed to the regulatory failure, such as limited resources and industry pressure, a key issue was that Federal Aviation Administration engineers did not fully grasp the risk profile of the Maneuvering Characteristics Augmentation System (MCAS), the software implicated in the crashes \citep{kitroeff2019roots, defazio2020boeing}. This lack of understanding of the underlying technology led to the software being erroneously categorised as low-risk, thereby bypassing more thorough scrutiny. 

Beyond the mere availability of expertise, it is also crucial to consider its sources. In order to operationalise policy goals, most AI regulatory regimes reference to standards \cite{pouget2023standards}. These are usually developed by standard-setting organisations, which in turn predominantly rely on contributions from the very industry it seeks to regulate. Such reliance is generally less of an issue in scenarios where the goals of the industry and regulators are closely aligned, like in ensuring the interoperability of electronic devices. However, the situation becomes more precarious when standard-setting organisations directly shape the regulatory obligations of developers. Put differently, regulatory requirements can have a direct impact on a developer's bottom line, creating the potential for regulatory capture \cite{dal2006capture}. Given this dynamic there's a need for greater involvement of the AI/ML research community (as well as other groups, such as representatives from civil society \cite{edwards2022regulating}) to ensure standards are well-informed and drive forward more robust and equitable regulatory practices \cite{cihon2019standards}. 

\subsubsection{Enforce Compliance}
After establishing policy priorities and converting them into tangible requirements, another crucial step in ensuring that rules will effectively bring about change lies in the regulator's capacity to enforce these regulations. While, in the first section, we mostly elaborated on the technical tools underlying enforcement, there is also a key question towards the people, infrastructure and expertise necessary to hold AI developers accountable \citep{scherer2015regulating}. Without a good understanding of the underlying practices in AI development, as well as the technology more generally, it will be difficult to detect violations and, for instance, scrutinise claims made by AI developers. Furthermore, narrowing information asymmetries not only increases the likelihood of identifying infringements but also strengthens overall compliance by presenting a more credible threat of detection and action \cite{ayres1992responsive, baldwin2008really}.

A concrete example where the expertise of the AI/ML research community could be crucial is in the context of the EU Liability Directive. While at the time of writing, this legislative act is still in the proposal stage, AI developers could be held liable for harms that are 1) foreseeable and 2) a likely result from a failure to meet ``duty of care'' \citep[Article 4]{euparliament2022ailiabilitydirective}. In essence, if a developer neglects established best practices and standards in AI system development, and this oversight leads to foreseeable harm, they could be liable \cite{li2022liability}. To investigate a claim, the current proposal would grant the European Commission powers to request technological evidence and documentation. However, determining misconduct might largely hinge on the regulator's ability to scrutinise the information provided by the industry and to ascertain whether the conditions of predictability (i.e., foreseeability of harm) and negligence of the duty of care (i.e., adherence to best practices) were met. Given that AI development is fundamentally grounded in practices from the AI/ML research community, the community is uniquely positioned to bridge some of the information gaps to hold AI developers accountable.

Alongside expertise housed within regulatory bodies and policymaking institutions, the effective enforcement of current AI regulations critically relies on the availability of independent, third-party auditing and testing services for external scrutiny \cite{hadfield2023regulatory, raji2022outsider, anderljung2023towards}. The responsibilities of such organisations are likely to include tasks such as red-teaming of AI systems on safety properties, or assessing the conformity with certain standards. While, from a policy perspective, there are still numerous uncertainties, such as the exact scope of audits and the processes for appointing and certifying auditors \cite{ojewale2024ai, guha2023ai}, a primary challenge, to begin with, is the limited availability of such organisations \cite{engler2023key}.

\subsubsection{Opportunities to work on AI Governance}

While there are numerous ways the AI/ML research community can contribute to AI governance efforts—more than we could exhaustively capture here—below we provide an overview of some key avenues that require varying levels of involvement:

\textbf{Part-time Opportunities}
    \begin{itemize}
        \item \textbf{Technical AI Governance Research}: This could involve developing new insights into open technical questions as outlined above, collaborating with interdisciplinary scholars in fields such as sociology, health, or law, or synthesising existing information for broader audiences, similar to the AI Index \cite{maslej2024index} or the International Scientific Report on AI Safety \cite{Bengio2024report}.
        \item \textbf{Policy Advisory Roles}: Another way to enhance government expertise without fully leaving academia is by taking on advisory roles, for instance by serving on the EU's AI scientific panel \cite{CouncilEU2024}, the UN's Multistakeholder Advisory Body on AI \cite{UNTechEnvoy2023}, or taking sabbaticals to contribute to policy development \cite{dworkin2024boost}. Other lower-commitment options that are open to all are participation in working groups or responding to requests for comments on governmental AI governance initiatives.
        \item \textbf{Standards Setting}: Researchers can shape AI governance and industry practices by contributing to national standards-setting bodies (e.g., NIST, CEN, CENELEC), which often detail regulatory provisions, or private standards organizations (e.g., ISO/IEC, IEEE) that are frequently referenced by national bodies and become common practice in the industry.
    \end{itemize}

\textbf{Full-time Opportunities}
    \begin{itemize}
        \item \textbf{Policy practice}: Recently, governments have placed a strong emphasis on recruiting technical talent, as evidenced by initiatives such as the National AI Talent Surge in the US \cite{whitehouse2024talent} or the focus on technical expertise in the first hiring round of the EU AI office \citeyearpar{EC2024hiring}. Furthermore, fellowship programs, including TechCongress \cite{TechCongress} and Horizon \cite{HorizonPublicService} in the US, can provide supported pathways into policy roles.
        \item \textbf{Think Tanks}: Conducting research at a think tank could involve translating technical insights into policy recommendations, mapping the landscape of AI developments, similar to Epoch's efforts in mapping AI trends \cite{sevilla2022compute}, or reviewing the effectiveness of current policies, such as Groves et. al \citeyearpar{groves2024auditing} work on reviewing algorithmic bias audit regime.
        \item \textbf{Regulatory Entrepreneurship}: Another opportunity to shape AI governance is by building organisations that provide the necessary AI accountability infrastructure, e.g., by developing privacy-enhancing technologies similar to OpenMined's work or providing evaluation and auditing services.
    \end{itemize}

We believe that technical AI governance offers many exciting opportunities. However, we also acknowledge that pursuing these paths may come with considerable costs for AI/ML researchers, such as significantly lower public sector salaries compared to industry role \cite{Bureau2023Wages, Fleming2023Wages} or slower progress towards tenure when undertaking additional responsibilities. 

Addressing these issues will require additional initiatives from other stakeholders in the AI governance ecosystem, such as universities, governments, and industry, to incentivize more technical AI governance work and interdisciplinary collaboration. While a thorough exploration of potential solutions exceeds this paper's scope, examples of such initiatives could include funding national interdisciplinary research programs, addressing pay disparities, ensuring legal protections for AI auditors \cite{longpre2024safe}, offering and promoting policy fellowships, or providing more opportunities for cross-disciplinary networking.

\section{Conclusion}\label{conclusion}


In this position paper we have argued for closer integration between AI researchers and policy makers in governments and other public-sector bodies. We surveyed policy documents published in the EU, US, and China, finding significant divergence between the current state of technical solutions and those assumed by, and necessary for enacting, proposed policy actions. In particular, we focused on technical requirements at the \textit{data}, \textit{compute}, \textit{model}, and \textit{deployment} level. Furthermore, we argued that government bodies will require more access to relevant technical expertise for \textit{informing}, \textit{operationalising}, and \textit{enforcing} governance and policy actions.

Based on these arguments we make the following two calls to action to the AI/ML research community:
\begin{enumerate}
    \item To prioritise research topics targeted at narrowing the gap between assumed and actual technical tools available for supporting governance efforts; and
    \item To work towards a closer integration with policy-makers, so as to ensure informed and effective governance of AI. 
\end{enumerate}

We look forward to following and supporting work towards these two aims, and plan on publishing a research agenda for the former as a next step in the near future.

\newpage
\section*{Impact Statement}
With this position paper, we seek to motivate the AI research community to actively engage with AI governance efforts, ultimately aiming to increase governance capacity to anticipate and respond to challenges posed by AI systems. We argue that AI researchers can contribute to this goal in two key ways: first, by creating and advancing the research necessary for effective oversight, and second, by providing their expertise to inform, operationalise and enforce policies. However, the technical research directions we propose only serve as illustrative examples rather than a comprehensive roadmap or a prioritisation of the most critical issues. Moreover, technical solutions and expertise are only one component of the AI governance toolbox; effective AI governance will require insights from other areas such as the social sciences, public engagement, and other interdisciplinary perspectives—areas that fall beyond the scope of this paper.

\bibliography{example_paper}
\bibliographystyle{icml2024}


\end{document}